# Status and overview of development of the Silicon Pixel Detector for the PHENIX Experiment at the BNL RHIC

**Ryo Ichimiya[a1], Nicole Apadula[b], Yasuyuki Akiba[a], Ermias Atomssa[c], Simon Chollet[c], Olivier Drapier[c], Hideto En'yo[a], Kohei Fujiwara[a], Franck Gastaldi[c], Raphael Granier de Cassagnac[c], Miki Kasai[d], Kazuyosi Kurita[d], Maki Kurosawa[a], Eric J. Mannel[e], Hiroaki Ohnishi[a], Yoshiyuki Onuki[a], Robert Pak[f], Charles Pancake[b], Michiko Sekimoto[g], Eugene Shafto[b], Walter Sondheim[h] and Atsushi Taketani[a]**

[a] *RIKEN Nishina Center for Accelerator-Based Science, 2-1, Hirosawa, Wako, Saitama 351-0198, Japan*
[b] *Stony Brook University, Stony Brook, NY 11794, U.S.A.*
[c] *LLR, Ecole Polytechnique, CNRS-IN2P3, 91128 PALAISEAU Cedex, France*
[d] *Rikkyo University, 3-34-1, Nishi-Ikebukuro, Toshima-ku, Tokyo, Japan*
[e] *Columbia University, New York, NY 10027, U.S.A.*
[f] *Brookhaven National Laboratory, Upton, NY 11973, U.S.A.*
[g] *High Energy Accelerator Research Organization, 1-1, Oho, Tsukuba, Ibaraki 305-0801, Japan*
[h] *Los Alamos National Laboratory, Los Alamos, NM 87545, U.S.A.*
*E-mail*: ryo@riken.jp

ABSTRACT: We have developed a silicon pixel detector to enhance the physics capabilities of the PHENIX experiment. This detector, consisting of two layers of sensors, will be installed around the beam pipe at the collision point and covers a pseudo-rapidity of $|\eta| < 1.2$ and an azimuth angle of $|\varphi| \sim 2\pi$. The detector uses 200 μm thick silicon sensors and readout chips developed for the ALICE experiment. In order to meet the PHENIX DAQ readout requirements, it is necessary to read out 4 readout chips in parallel. The physics goals of PHENIX require that radiation thickness of the detector be minimized. To meet these criteria, the detector has been designed and developed. In this paper, we report the current status of the development, especially the development of the low-mass readout bus and the front-end readout electronics.

[1] Corresponding author

## Contents

## 1. Introduction

The Pioneering High Energy Nuclear Interaction eXperiment (PHENIX) at the Relativistic Heavy Ion Collider (RHIC) at the Brookhaven National Laboratory (BNL), which has been exploring both spin structure of the nucleon utilizing polarized proton-proton collisions and characteristics of the Quark Gluon Plasma created in heavy ion collisions, is being upgraded with a silicon vertex tracker (VTX) to gain precise information near the collision point. The primary purpose of the VTX is to carry out precise measurements of heavy-quark production (charm and beauty) in $A + A$, $p(d) + A$ and polarized $p + p$ collisions. These are essential measurements for future heavy-ion and spin-physics programs at RHIC. The VTX must be able to distinguish heavy quarks from light quarks by detecting a displaced vertex due to the longer lifetime of heavy quarks, whose $c\tau$ range from 100 μm (charm) to 400 μm (bottom). In addition, the precise tracking capabilities in high charged particle densities $dN / d\eta \sim 700$ (at $\eta = 0$), of the VTX will provide significant improvements to other physics measurements.

## 2. Silicon pixel detector

The VTX detector consists of two inner layers of silicon pixel detectors with radii of 25 and 50 mm and two outer layers of silicon strip detectors with radii of 100 and 140 mm, as shown in Figure 1. The detector covers a pseudo-rapidity of $| \eta | < 1.2$ and $\Delta\varphi \sim 2\pi$. The silicon strip detector is a single-sided, DC-coupled, two-dimensional silicon detector, which was developed at BNL[1]. The specifications of the VTX pixel detectors are summarized in Table 1[2–3].

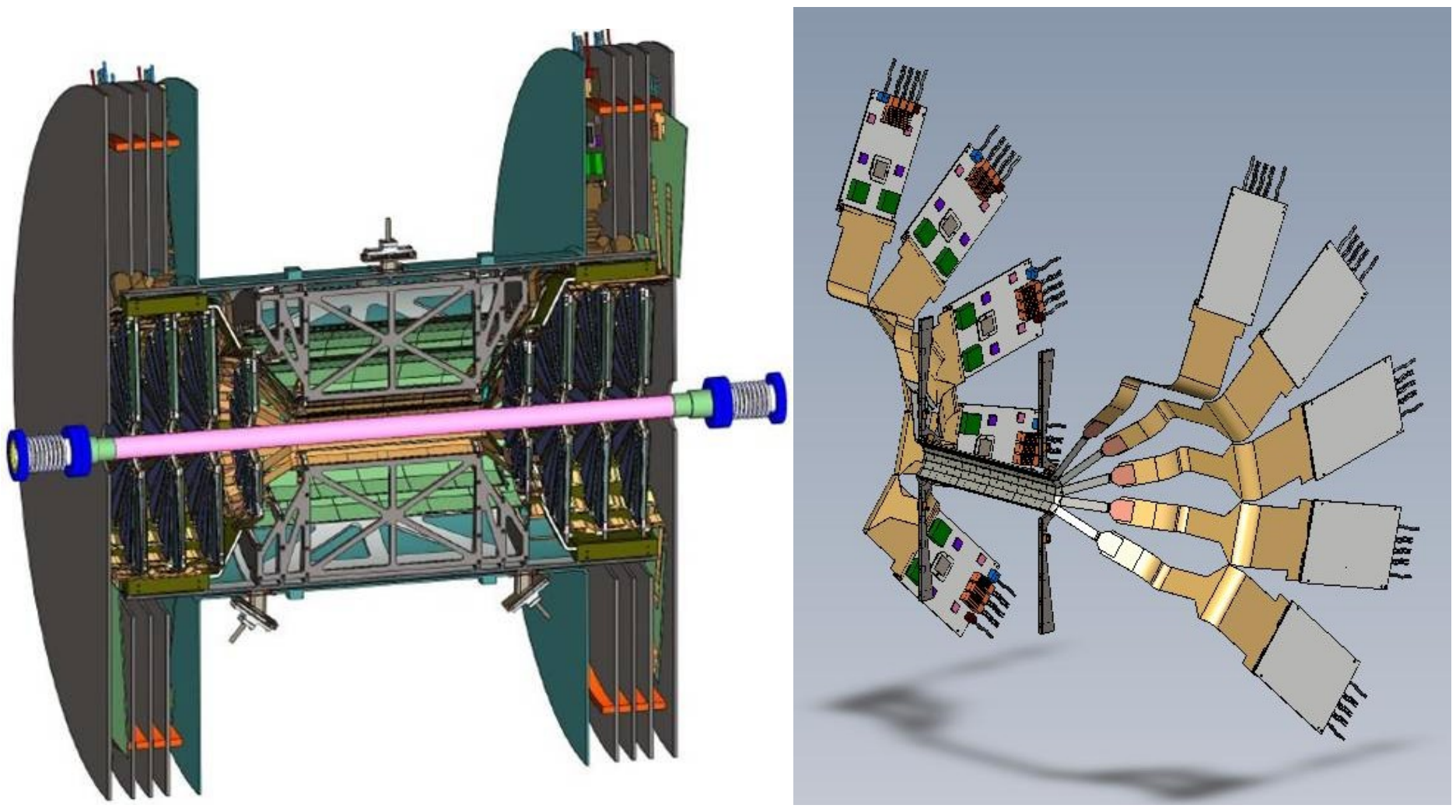

**Figure 1 Three-dimensional drawing of the VTX with the beam pipe (Left). The VTX sensor consists of two inner layers of silicon pixel detectors (25 mm, 50 mm) and two outer layers of silicon strip detectors (100 mm, 140 mm). The five pixel ladders with the both sides of the bus extenders and the SPIRO modules are displayed (Right). The readout bus is located at the inner barrel region and the SPIRO modules are located at the outer region.**

| VTX | Layer | R1 | R2 |
|---|---|---|---|
| Geometrical dimensions | R (mm) | 25 | 50 |
| | Δz (mm) | 218 | |
| | Area ($mm^2$) | 28,000 | 56,000 |
| | Number of Ladders | 10 | 20 |
| | Readout channels | 1,310,720 | 2,620,440 |
| Sensor hybrid | Sensor size | 128 mm × 13.6 mm | |
| | ALICE1/LHCb chips | 15.6 mm × 13.5 mm (× 4) | |
| | Pixel size | 50 µm × 425 µm | |
| | Readout chips | 160 | 320 |
| Radiation length per layer ($X/X_0$) | Sensor | 0.21% | |
| | Readout | 0.16% | |
| | Bus | 0.22% | |
| | Support and cooling | 0.67% | |
| | Total | 1.26% | |

**Table 1 Main parameters of silicon pixel layers.**

The technology used for the two inner layers of pixel detectors is based on the ALICE ITS detector developed at CERN. A 200-µm-thick silicon sensor fabricated by CANBERRA is a modified version of the ALICE sensor[4], that holds 4 (readout chip) × 32 (column) × 256 (row) pixels, each with an active area of 50 × 425 $\mu m^2$. We used the same readout chip[5] as ALICE ITS uses. The sensor is bump bonded to four matching readout chips, each having a thickness of 150 µm. Each readout chip has 32 × 256 amplifier-discriminators for individual channels. The data is read out in a pipeline fashion. Four readout chips mounted on a sensor using bump-bonding technology form a sensor hybrid. Two sensor hybrids are glued to a pixel readout bus made of Cu-Al-Polyimide, which is supported by a carbon fiber structure. The pixel readout bus is connected to the Silicon Pixel Interface Read-Out (SPIRO) module via a bus extender, which provides all service voltages, control and timing signals, and reads out the pixel data. The SPIRO module transmits the data to the Front End Module (FEM) via serial optical links. The FEM interprets and sends the data to the PHENIX data acquisition system (DAQ). The Bus extender is a Flexible Printed Circuit board (FPC) which feeds the data from the readout bus in the inner barrel to the SPIRO module in the outer region (See Figure 1(Right)).

A pixel readout bus, a bus extender, and one SPIRO module form a half ladder, the basic building unit of the pixel detector. Two half ladders form a full ladder, which spans approximately 22 cm in the beam direction. Ten ladders in the first layer and 20 ladders in the second layer together cover almost the entire azimuthally angle. The combined materials of silicon sensor, readout chip, readout bus, and mechanical support including the cooling structure add up to about 1.26% of a radiation length for one pixel layer.

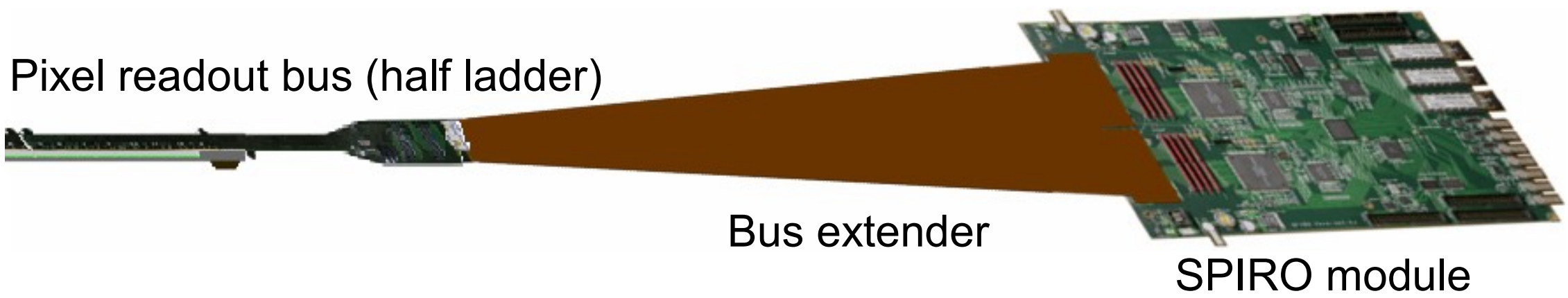

**Figure 2 A half ladder with bus extender and SPIRO module. The SPIRO module is located at the outer region, and the bus extender connects the pixel readout bus to the SPIRO module, as shown in Figure 1.**

## 2.1 Pixel Readout Bus

To meet the readout speed requirements of PHENIX, we adopted parallel readout scheme (See Figure 5 and Section 2.2). Therefore the pixel readout bus is required to have 188 signal lines, 250mm long, within a 13.9mm width. Moreover, in order to minimize the multiple scattering and photon conversion in the structure, the readout bus must be as thin in material as possible. To satisfy these criteria, we selected the FPC technology to fabricate the bus. The development of the pixel readout bus was challenging business for fabricating a high density (60 μm pitch) *and* long length (250 mm) bus.

We have developed a new copper-aluminium-polyimide FPC that has 188 signal lines with 0.22% $X_0$ in radiation length. The bus has four copper signal layers of 3 μm thickness and two aluminum power layers of 50 μm thickness. For the signal layers, we used 30 μm line width and spacing. In order to achieve a high yield in the manufacturing stage, we had to optimize the wiring pattern, through hole via plating and fabrication process.

We have done a signal transmission simulation of the pixel readout bus using the HSPICE simulator to evaluate the performance of the pixel bus for resistance, transmitted waveforms, and cross talks on the neighbouring lines[6]. A transmission line model with actual cross section and length of line was used for the simulation input. In addition, realistic models of the transmitter and receiver circuits were used. The results of the simulation were compared to the propagation characteristics, which were measured using a prototype half ladder. The simulation results and measured values were in good agreement and the crosstalk level was sufficiently low for the stable operation of the half ladder. We tuned the pull-up resistor values using this simulation method and concluded a 220 Ω termination resistor gives the best propagation characteristics for the half ladder.

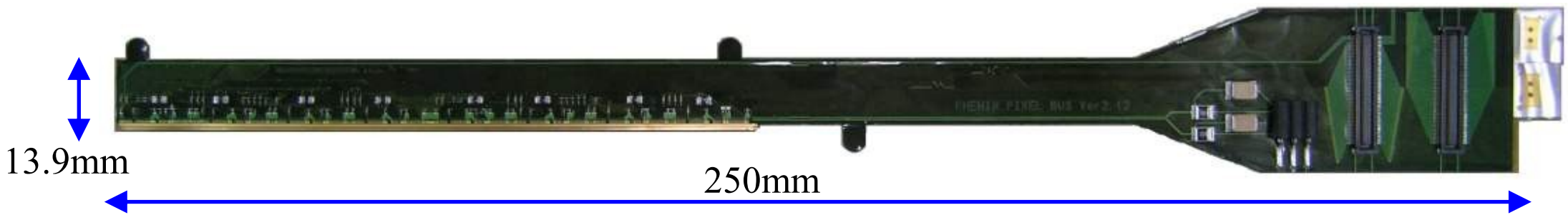

**Figure 3 Photograph of pre-production pixel readout bus. This bus is fabricated using fine-pitch (60 µm) FPC technology, and it has a length of 250 mm and a width of 13.9 mm.**

## 2.2 SPIRO module

As mentioned above, the SPIRO module is the frontend electronics for the half-ladders. It consists of two Digital Pilot ASICs, two Analogue Pilot ASICs, a control FPGA, two Gigabit Optical Link (GOL) high-speed serializer ASICs, which were developed at CERN and three optical transceivers as shown in Figure 4. Figure 5 shows the readout diagram from the pixel bus through the SPIRO module to the PHENIX DAQ.

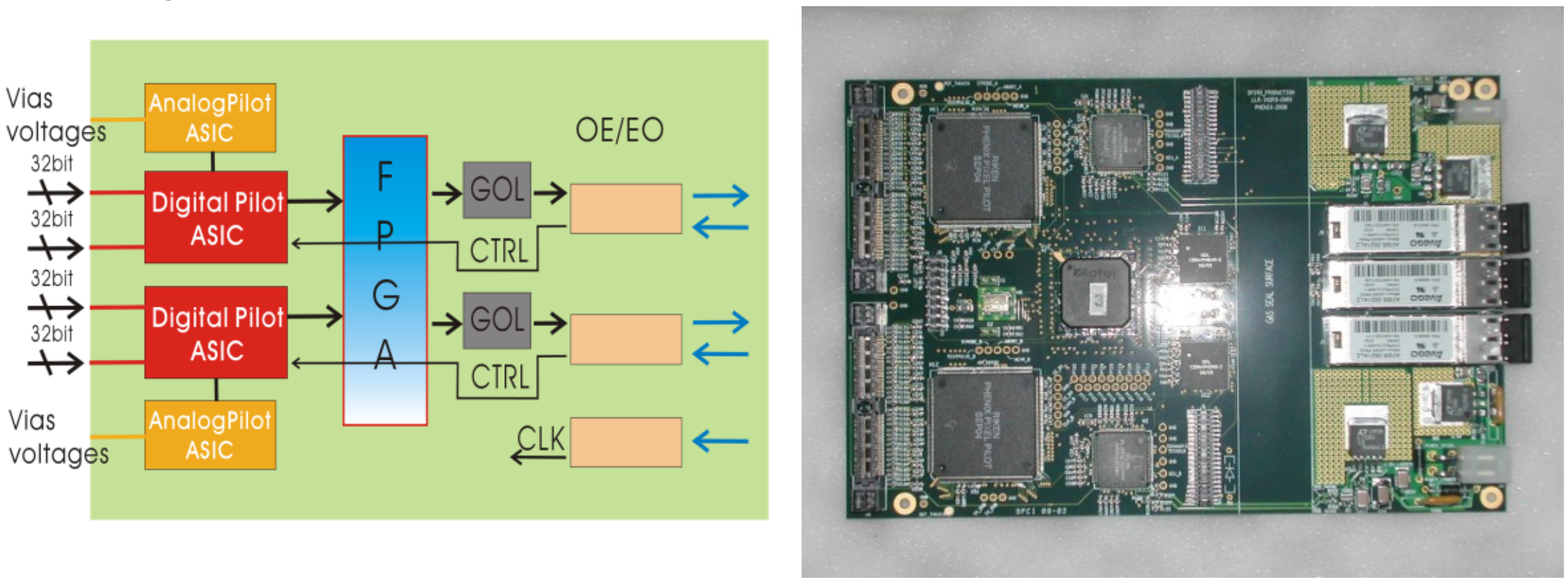

**Figure 4 Block diagram of SPIRO module (Left); photograph of SPIRO module (Right).**

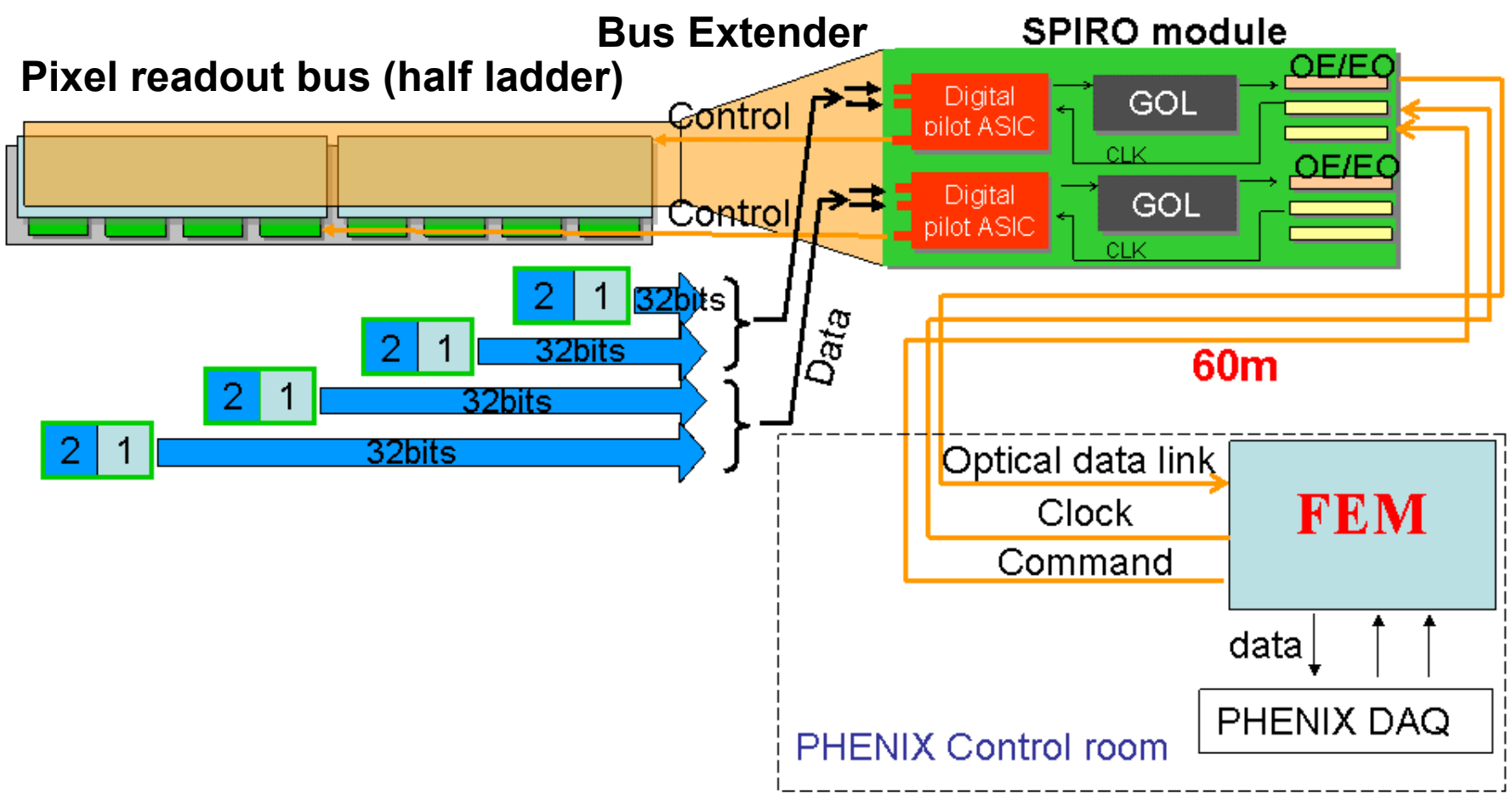

**Figure 5 Readout diagram from the pixel readout bus through the SPIRO module to the PHENIX DAQ.**

The SPIRO module performs the following functions:

- Transmitting control signals to the half-ladder.
- Multiplexing data from the half-ladder and transmitting them to the FEM via optical links.
- Providing slow control of the pixel readout chips and the SPIRO module itself.

The core components of the SPIRO module are the Digital Pilot ASICs. Each of the two pixel readout chip pairs presents 256 × 2 sequential words of data to a 32-bit bus synchronously at a beam collision clock frequency (BCO) of 9.4 MHz. To meet the readout speed requirements of PHENIX, four readout chips are read out in parallel. To optimize the readout speed and space factor, the PHENIX Digital Pilot ASIC with twice the number of input channels was developed using the same design rules and radiation tolerant technology as the original ALICE Digital Pilot ASIC[7]. The PHENIX Digital Pilot ASIC with 2 × 32 inputs can simultaneously read two 32-bit words from two pixel readout chips. Each 32-bit input handles the output from a pair of chips, which represents 512 sequential words of pixel data. Each of the two Analogue Pilot ASICs provides several reference voltages using on-chip Digital-to-Analogue Converters and monitors the voltages and currents of four pixel readout chips using an on-chip Analogue-to-Digital Converter. We use the Analogue Pilot ASICs without any modifications. The FPGA transfers the data from the PHENIX global clock (4 × BCO) domain to a low-jitter local clock (40 MHz) domain using an on-chip First-In First-Out memory. Two GOL ASICs serialize the 32-bit-wide data into a 1.6 Gbps serial string.

## 3. Beam test

A full chain test was conducted using a telescope with three layers of pre-production pixel ladders, bus extenders, SPIRO modules, a FEM and a PHENIX DAQ system to demonstrate the performance of the system. This test was conducted using a 120 GeV proton beam at the MTEST Facility at Fermi National Accelerator Laboratory (FNAL) in August 2008, FNAL T-984.

Figure 6 shows the beam test setup. A strip detector is also installed to the setup and the data are taken independently. Three pixel half ladders are stacked along the beam direction. Each half ladder is connected to a SPIRO module. All the SPIRO modules are controlled by a single FEM via optical fibres. The trigger required all 3 layers having a hit in coincidence with scintillation trigger counters, which were located upstream and downstream of the telescope. When the FEM received the trigger signal, it transmits a level-1 trigger to all the half ladders. The pixel hit information is then collected by the PHENIX DAQ system through the SPIRO modules and the FEM.

All systems were operated as designed and the hit data were recorded by the PHENIX DAQ system at a trigger rate of typically 34–38Hz. This trigger rate is considerably lower than that of the actual PHENIX (up to 10 kHz) and the number of input channels to the DAQ system is very small. Otherwise all other experimental conditions are identical to those of the PHENIX experiment. A typical beam event is shown in Figure 7. The three figures on the left show the hit positions on each layer after clustering of hits. The figure on the right represents the three-dimensional hit positions and the reconstructed track after geometrical alignment was performed.

Figure 8 shows the residual distribution of the first layer (Layer 1 in Figure 7) after geometrical alignment. The pixel size is 50 μm × 425 μm. The z direction is along the 425-μm side of the pixel, and the φ direction is along the 50-μm side of the pixel. The pixel sensor is expected to have resolutions of 123 μm (=425/$\sqrt{12}$ ) and 14 μm (=50/$\sqrt{12}$ ) in the z and φ directions, respectively. We determined the intrinsic position resolution of the sensor hybrid to be 109 ± 37 μm and 13 ± 5 μm in the z and φ directions, respectively. The error is the statistical error not including the systematic error. They agreed with the expected values well.

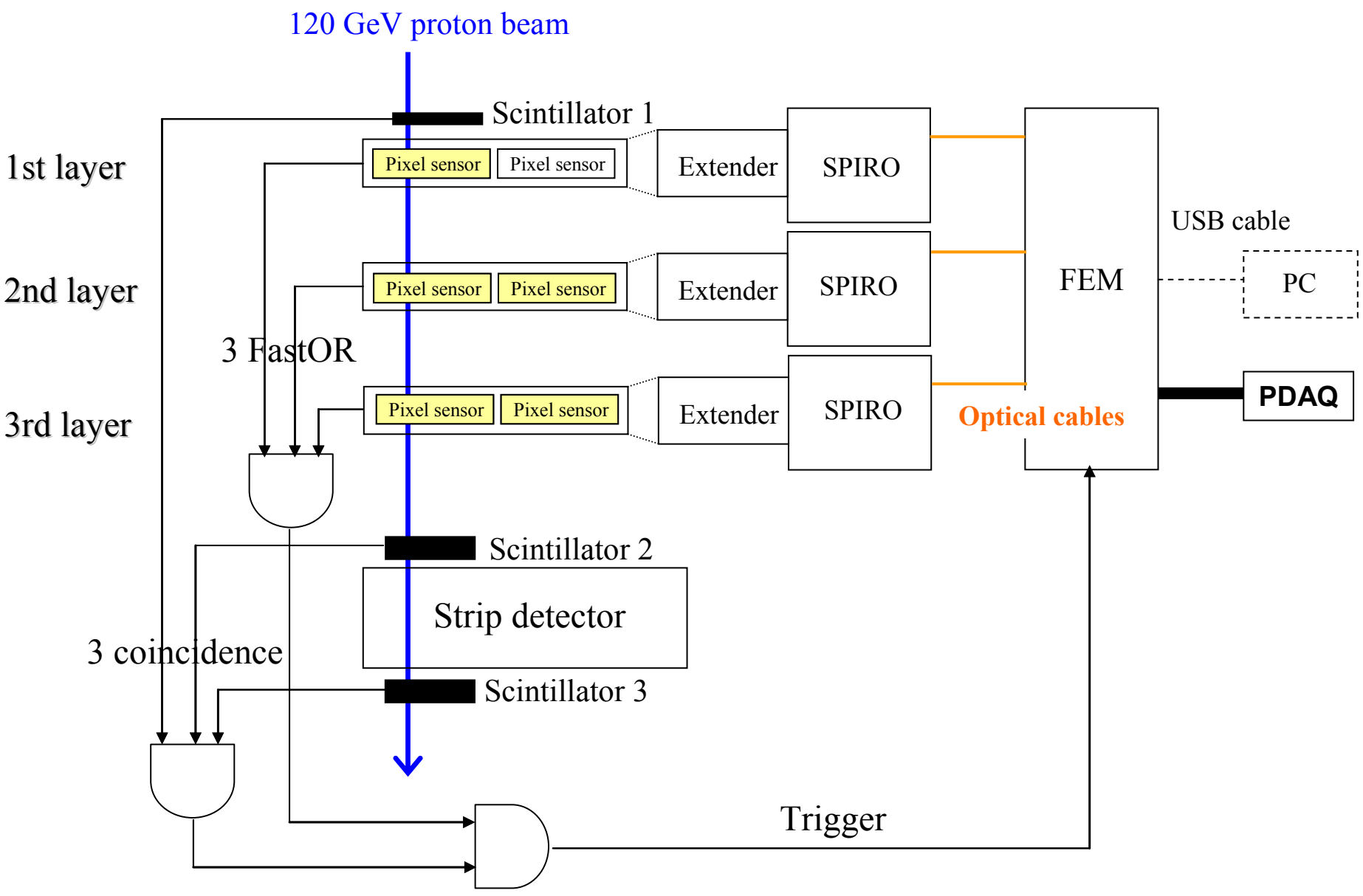

**Figure 6 Setup of beam test. The proton beam passes through each sensor hybrid from top to bottom. Three pixel half ladder layers are stacked along the beam direction.**

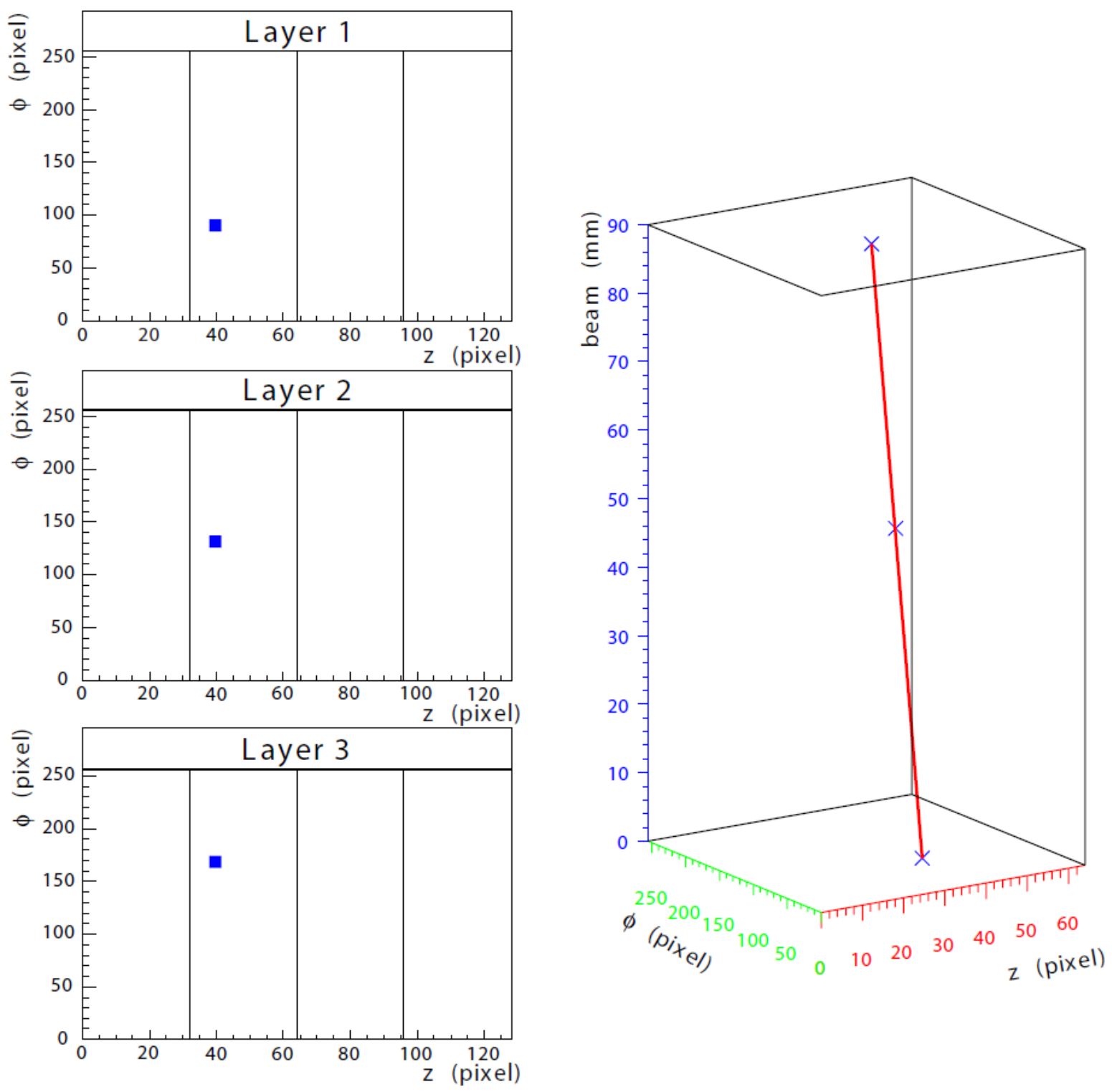

**Figure 7 Hit positions at each pixel layer after clustering of consecutive hits (Left) and three-dimensional hit positions and reconstructed track (Right).**

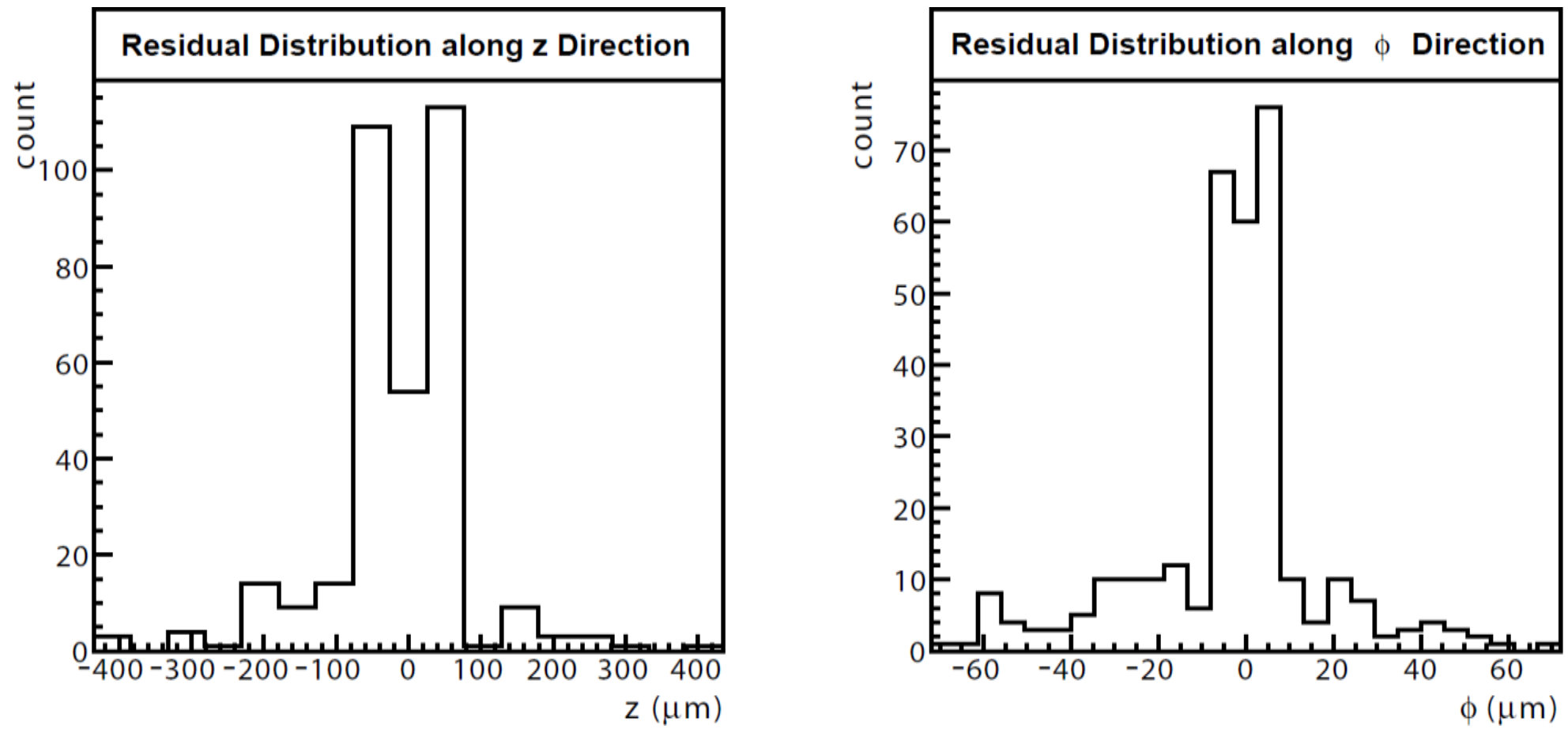

**Figure 8 Residual distributions of Layer 1 along z and φ directions. Pixel size is 50 μm × 425 μm; z direction is along the 425-μm side of the pixel and φ direction is along the 50-μm side of the pixel.**

## 4. Summary

We have developed a silicon pixel detector for the PHENIX upgrade, using a sensor and a readout chip developed for the ALICE experiment at CERN. Based on the original ALICE pixel detector system, we have developed the pixel readout bus, bus extender, SPIRO module, and FEM module. We have succeeded in developing a high-density (188 traces in a width of 13.9 mm) *and* thin (radiation length of 0.22% $X_0$) pixel readout bus. A full chain test was conducted on the pixel detector system using pre-production pixel ladders, frontend electronics, and the PHENIX DAQ system at the MTest Facility at FNAL in August 2008, to demonstrate the system performance.

## Acknowledgments

We would like to thank everyone who helped us in this study, especially our collaborators in the CERN micro electronics group and the ALICE pixel detector group. We thank Mr. Bista Deepak and Mr. Jumpei Kanaya for their efforts on chip QA. We also wish to thank the staff of the FNAL MTest Facility and Accelerator Division.